\documentclass[aps,prl,superscriptaddress,twocolumn,longbibliography]{revtex4-1}
\usepackage{bbm}
\usepackage{mathrsfs}
\usepackage{amsmath}
\usepackage{amsfonts}
\usepackage[colorlinks=true,linkcolor=blue,urlcolor=blue,citecolor=blue,anchorcolor=blue]{hyperref}
\usepackage{graphicx,epstopdf}
\usepackage{subfigure}
\usepackage{epsfig}
\usepackage{dcolumn}
\usepackage{bm}
\usepackage{color}
\usepackage{natbib}
\usepackage{amssymb}
\usepackage{xcolor}
\usepackage{braket}
\usepackage{ulem}
\usepackage{float}%H
\usepackage{braket}
\usepackage{changes}
\allowdisplaybreaks[4]

\begin{document}
	
\title{Phase transition and multistability in Dicke dimer}
\author{Yilun Xu}
\thanks{These authors contributed equally to this work.}
\affiliation{State Key Laboratory for Mesoscopic Physics, School of Physics, Frontiers Science Center for Nano-optoelectronics, Peking University, Beijing 100871, China}
\affiliation{Beijing Academy of Quantum Information Sciences, Beijing 100193, China}
\affiliation{Department of Physics and Astronomy, Rice University, Houston, Texas 77251-1892, USA}
\author{Feng-Xiao Sun}
\thanks{These authors contributed equally to this work.}
\affiliation{State Key Laboratory for Mesoscopic Physics, School of Physics, Frontiers Science Center for Nano-optoelectronics, Peking University, Beijing 100871, China}
\author{Wei Zhang}
\email{wzhangl@ruc.edu.cn}
\address{Department of Physics and Key Laboratory of Quantum State Construction and Manipulation (Ministry of Education), Renmin University of China, Beijing 100872, China}
\address{Beijing Academy of Quantum Information Sciences, Beijing 100193, China}
\author{Qiongyi He}
\email{qiongyihe@pku.edu.cn}
\affiliation{State Key Laboratory for Mesoscopic Physics, School of Physics, Frontiers Science Center for Nano-optoelectronics, Peking University, Beijing 100871, China}
\affiliation{Collaborative Innovation Center of Extreme Optics, Shanxi University, Taiyuan 030006, China}
\affiliation{Peking University Yangtze Delta Institute of Optoelectronics, Nantong 226010, China}
\affiliation{Hefei National Laboratory, Hefei 230088, China}
\author{Han Pu}
\email{hpu@rice.edu}
\affiliation{Department of Physics and Astronomy, Rice University, Houston, Texas 77251-1892, USA}

\begin{abstract}
The exotic phase transitions and multistabilities in atom-cavity coupled systems have attracted tremendous interests recently. In this work, we investigate the effect of photon hopping between two Dicke cavities, which induces rich quantum phases for steady states and dynamic process. Starting from a generic dimer system where the two cavities are not necessarily identical, we analytically prove all possible steady-state phases, which are confirmed by numerical calculations. We then focus on the special case with two identical cavities, where all the steady states are confirmed by exact solutions. We show that photon hopping is a convenient and powerful tool to manipulate the quantum phases and induce multistable behavior in this system.  
\end{abstract} 
	
\maketitle
\textit{Introduction.} 
%To begin with our work, we want to raise a daily question: how would it be like if we drop a block of ice into a cup of water? Obviously, we have three possible final states: a cup of water, a cup of ice or the mixture of water and ice. Such a nature phenomenon in our daily life reveal how the two different phases (liquid and solid) affect each other. But when it comes towards quantum world, what will happen then?
Quantum phase transition (QPT) is a phenomenon describing the sudden change of quantum properties caused by the continuous change of system parameter(s). With the recent rapid progress in experimental techniques on light-matter coupling~\cite{RN102,RN83,RN84,RN86,RN87,RN88,doi:10.1126/science.abd4385}, QPTs have been observed in cold atomic systems coupled with cavities~\cite{RN102, doi:10.1126/science.abd4385}, trapped ions~\cite{RN83,PhysRevLett.118.073001}, superconducting circuits~\cite{RN84}, and so on. Although these physical platforms look very different from each other, many of them can be described by the Dicke model or one of its variants~\cite{HEPP1973360,PhysRevA.7.831,PhysRevLett.90.044101,PhysRevE.67.066203}. The Dicke model, describing an ensemble of two-level atoms interacting with a cavity field, is one of the most fascinating quantum optical models supporting QPT between a normal phase (NP) and a superradiant phase (SRP) in the thermodynamic limit which can be reached as the number of atoms $N\rightarrow\infty$.
%or equivalently, $\Omega/\omega\rightarrow\infty$ in single atom-cavity coupled system~\cite{PhysRevLett.115.180404,PhysRevLett.117.123602}. Here, $\Omega$ and $\omega$ stand for the transition frequency of the two level atom and the cavity field frequency respectively. 

More recently, great attention has been paid to the investigations of quantum phases and QPT in multi-cavity  systems~\cite{PhysRevLett.115.180404,PhysRevA.101.063843,PhysRevLett.127.063602,PhysRevLett.129.183602,PhysRevLett.117.123602,PhysRevLett.128.163601,PhysRevResearch.5.L042016}, nonreciprocal effect induced phase transition~\cite{PhysRevLett.122.193605,PhysRevLett.131.113602}, and multistabilities and nonequilibrium dynamics of quantum phases~\cite{PhysRevLett.120.183603,PhysRevLett.124.073602,PhysRevLett.112.173601,PhysRevLett.132.073602,PhysRevLett.105.043001}. Among them, the photon hopping effect~\cite{PhysRevA.101.063843,PhysRevLett.128.163601,PhysRevLett.117.123602,PhysRevLett.127.063602,PhysRevLett.128.163601,PhysRevResearch.5.L042016} in multi-cavity systems plays an essential role in manipulating novel quantum phases. However, a systematic investigation of hopping induced nonequilibrium dynamics and multistabilities is still lacking. Furthermore, in most studies of multi-cavity systems, it has been conveniently assumed that the cavities are all identical, i.e., they are characterized by the same set of parameters. This assumption greatly simplifies calculation, in the expense of a loss of generality.

In this Letter, we investigate the steady-state QPT of a Dicke dimer system, where two Dicke cavities (which, for the general case, are not necessarily identical) are coupled through photon hopping. In the absence of such photon hopping, we have two isolated Dicke cavities, and the combined phases are: both cavities in the normal phase (NP\&NP), one cavity in normal and the other in superradiant phase (NP\&SRP), and both cavities in the superradiant phase (SRP\&SRP). Through a rigorous and analytic proof, we show that as long as the photon hopping is present, both cavities must be in the same phases, i.e., (NP\&SRP) cannot occur. We obtain the phase diagram and also numerically identify a multistable regime where different stable superradiant states exist. To gain further analytic insights, we then consider two identical cavities and show that the multistable regime consists of a symmetric superradiant phase (SSRP) and an anti-symmetric superradiant phase (ASRP). The dynamic evolution of various quantum phases are studied in both open and closed systems. When two isolated systems are coupled together, the coupling between them can often leads to new phenomena. The Dicke dimer offers an ideal platform to study such coupling in a controlled manner.  

\textit{Model of the Dicke dimer.} We will concentrate on investigating the two coupled Dicke cavities and its phase diagram, in which the photon hopping plays an important role. The Hamiltonian of the system can be expressed as $H_{\rm sys}=\sum_{j=1,2}H_j^{\rm Dicke}+H_{\rm XX}$.
%\begin{align}
%	\label{XX Hamiltonian}
%	H&=\sum_{j=1,2}H_j^{Dicke}+H_{XX}\\
%	H_j^{Dicke}&=\omega_ca_j^\dagger a_j+\omega_a S_j^z+\dfrac{2\lambda_x}{\sqrt{N}}S_j^x(a_j+a_j^\dagger)\\
%	H_{XX}&=J(a_1+a_1^\dagger)(a_2+a_2^\dagger),
%\end{align}
The Dicke Hamiltonian in cavity $j=1,2$ reads $H_j^{\rm Dicke}=\omega_{cj}a_j^\dagger a_j+\omega_{aj} S_j^z+{2\lambda_j}S_j^x(a_j+a_j^\dagger)/{\sqrt{N}}$, where $a_{j}$ is the photon annihilation operator of the single cavity mode, and $S_{j}^{x,y,z}\equiv\sum_{i=1}^{N}s_{i,j}^{x,y,z}$ are collective spin operators of the atomic ensemble in the cavity $j$ which contains $N$ two-level atoms, with $s_{i,j}^{x,y,z}$ standing for the single spin operators of the $i$th atom in cavity $j$. The optical frequency and the atomic transition frequency are denoted by $\omega_{cj}$ and $\omega_{aj}$, respectively, and $\lambda_j$ stands for the atom-cavity coupling strength in cavity $j$. For each isolated cavity, it is in the normal (superradiant) phase if $\lambda_j < \lambda_{cj}$ ($\lambda_j > \lambda_{cj}$), where the critical coupling strength is given by~\cite{PhysRevLett.120.183603}
\begin{equation}
   \lambda_{cj}= \frac{1}{2}\sqrt{\omega_{aj}(\omega_{cj}+{\kappa_j^2}/{\omega_{cj}})}\,
\end{equation}
with $\kappa_j$ the respective cavity decay rate.
%These parameters are set the same for both cavities, because we focus on a symmetric Dicke dimer where the Dicke cavities are assumed identical. 
The photon hopping between the cavities is described by $H_{\rm XX}=J(a_1+a_1^\dagger)(a_2+a_2^\dagger)$ with hopping amplitude $J$~\cite{PhysRevA.101.063843,PhysRevLett.113.093602,PhysRevLett.128.160504}. In this work, we will take $J$ to be real and positive.

For a given operator $\mathcal{O}$, its dynamics can be described by the master equation ${d\mathcal{O}}/{dt}=i[H_{\rm sys},\mathcal{O}]+\mathcal{L}[\mathcal{O}]$, where the Lindblad superoperator arising from the cavity decay reads $\mathcal{L}[\mathcal{O}]=\sum_{j=1,2}\kappa_j(2a_j^\dagger\mathcal{O}a_j-\mathcal{O}a_j^\dagger a_j-a_j^\dagger a_j\mathcal{O})$. Applying the mean-field approximation, we substitute the operators by their corresponding expectation values as $\langle a_{j}\rangle=\sqrt{N}\gamma_{j}$ and $\langle S_{j}^x\rangle=NX_{j}$, $\langle S_{j}^y\rangle=NY_{j}$, $\langle S_{j}^z\rangle=NZ_{j}$,
%Our following result is suitable for any atom number $N$. But when it refers to the quantum fluctuation, the thermodynamic limit $N\rightarrow\infty$ will make a difference to suppressing the quantum fluctuation thus making our analytical solutions more accurate.
and obtain 
\begin{align}
\label{eq:EOM}
%	\begin{cases}
		\frac{d{\rm Re}\gamma_{j}}{dt}&=-\kappa_{j} {\rm Re}\gamma_{j}+\omega_{cj}{\rm Im}\gamma_{j},\nonumber\\
		\frac{d{\rm Im}\gamma_{j}}{dt}&=-\omega_{cj}{\rm Re}\gamma_{j}-\kappa_{j} {\rm Im}\gamma_{j}-2J{\rm Re}\gamma_{3-j} -2\lambda_{j} X_{j},\nonumber\\
		\frac{dX_{j}}{dt}&=-\omega_{aj}Y_{j},\nonumber\\
		\frac{dY_{j}}{dt}&=\omega_{aj}X_{j}-4\lambda_{j}Z_{j}{\rm Re}\gamma_{j},\nonumber\\
		\frac{dZ_{j}}{dt}&=4\lambda_{j}Y_{j}{\rm Re}\gamma_{j}.
%	\end{cases}.
\end{align}
Here, we have assumed that the atomic and the photonic operators are uncorrelated, an approximation valid and well tested in the thermodynamic limit. Considering the indistinguishability of the atoms in the same cavity, the total spin conservation law $\langle S_{j}^x\rangle^2+\langle S_{j}^y\rangle^2+\langle S_{j}^z\rangle^2={N^2}/{4}$ is valid. Thus the last equation in Eq.~(\ref{eq:EOM}) is redundant. 

%\begin{figure}[tb]
%	\centering
%	\includegraphics[width=0.212\textwidth]{asymmetry/phase-trans-asy}
%	\includegraphics[width=0.265\textwidth]{asymmetry/tot6}
%	\caption{(a) The possible phase transitions of the asymmetric Dicke dimer with $J_i=0\rightarrow J_f\neq0$ indicated by the solid arrows. (b) The critical hopping $J_c$ against the asymmetric atom cavity coupling strengths $\lambda_{1}$ and $\lambda_{2}$, with $\kappa_1=\kappa_2=0.2$ and $\omega_{a1}=\omega_{a2}=\omega_{c1}=\omega_{c2}=1$.}
%	\label{table}
%\end{figure}

\begin{figure}[tb]
	\centering
	\hspace*{-0.2cm}
	\includegraphics[width=0.5\textwidth]{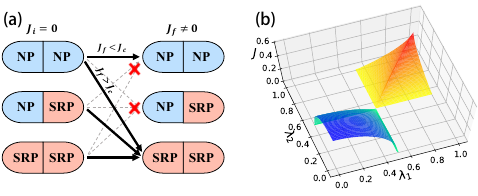}
	\caption{(a) The possible phase transitions of the Dicke dimer with $J_i=0\rightarrow J_f\neq0$ indicated by the solid arrows. (b) The critical hopping $J_c$ (blue) and the multistable boundary (orange) against the atom-cavity coupling strengths $\lambda_{1}$ and $\lambda_{2}$, with $\kappa_1=\kappa_2=0.2$ and $\omega_{a1}=\omega_{a2}=\omega_{c1}=\omega_{c2}=1$.}
	\label{table}
\end{figure}

\textit{Steady-state phase diagram.} Let us consider the following scenario. Initially, we turn off the photon hopping by setting $J_i=0$ and each cavity reaches the steady state. And then we turn on the photon hopping with $J_f=J \neq 0$ and let the system evolve to a new steady state in the long time limit. To study the dynamical evolution, we define an $8$-dimensional vector of expectation values as $\Psi_{\rm tot}^T\equiv[\Psi^T_1,\Psi^T_2]$, where $\Psi_{j}\equiv[{\rm Re}\gamma_{j},{\rm Im}\gamma_{j},X_{j},Y_{j}]^T$. The dynamic equations are given by ${d\Psi_{\rm tot}}/{dt}=\mathcal{D}\Psi_{\rm tot}$ where the $8\times8$ coefficients matrix $\mathcal{D}$ can be straightforwardly obtained from Eq.~(\ref{eq:EOM}) and its explicit form can be found in the Supplemental Materials (SM)~\cite{supplementary}. 

Depending on whether $\lambda_j$ is smaller or larger than the critical coupling strength $\lambda_{cj}$, the initial state of the dimer can be labeled by one of the following: (NP\&NP), (NP\&SRP) and (SRP\&SRP). The new steady state after turning on the photon hopping can in principle be obtained by taking ${d\Psi_{\rm tot}}/{dt}=0$. However, for two generic non-identical cavities, analytic solutions are intractable. Nevertheless, we can analytically prove the following statements (for details, see SM~\cite{supplementary}):
\begin{itemize}
	\item[(i)] In the final steady state, both cavities will always be in the same phase, i.e., either (SRP\&SRP) or (NP\&NP). 
	\item[(ii)] The initial (NP\&NP) at $J_i=0$, which occurs when $\lambda_j<\lambda_{cj}$ for both $j=1,2$, will be evolved into (SRP\&SRP) if the final hopping rate $J>J_c$, and remain in (NP\&NP) if $J<J_c$, where the critical photon hopping rate is given by 
 \begin{equation} 
 J_c=2\sqrt{\left(\lambda_{c1}^2-\lambda_1^2\right) \left(\lambda_{c2}^2-\lambda_2^2\right) /(\omega_{a1} \omega_{a2})  }\,.
 \end{equation}
	\item[(iii)] If initially any one of the cavities is in SRP, then the final state will be (SRP\&SRP). 
\end{itemize}
These qualitative conclusions are summarized in Fig.~\ref{table}(a). In Fig.~\ref{table}(b), we present a steady-state phase diagram in the parameter space spanned by $J$, $\lambda_1$ and $\lambda_2$ while fixing all other parameters as indicated in the figure caption. The blue surface represents $J_c$, the region under which is (NP\&NP). In all the other region, we have (SRP\&SRP). The region under the orange surface (obtained numerically) is a multistable region, where one can find four stable states, which are connected to the four degenerate SRP states in the initial state with $J_i=0$. These four degenerate SRP states result from the spontaneous breaking of the $Z_2\times Z_2$ symmetry of individual cavities. In the rest of the (SRP\&SRP), one pair of the states become dynamically unstable and only the other pair of states survive, corresponding to the spontaneous breaking of the total $Z_2$ symmetry of the dimer~\cite{Z2_symmetry}. We also want to remark that the system would not be stable for sufficiently large $J$. For the generic case here, it is difficult to obtain analytically the threshold of $J$ beyond which the system is unstable. However, for two symmetric cavities to be discussed later, we are able to obtain this threshold and, furthermore, show how additional nonlinear terms can stabilize this regime.

%\begin{figure}[tb]
%	\centering
%	\subfigure{\includegraphics[width=0.22\textwidth]{asymmetry/0.25,0.35,J=0.4.pdf}}
%	\hspace*{-0.3cm}
%	\subfigure{\includegraphics[width=0.22\textwidth]{asymmetry/0.45,0.55,J=0.2}}\\
%	\vspace*{-0.52cm}
%	\subfigure{\includegraphics[width=0.22\textwidth]{asymmetry/0.7,0.8,J=0.1}}
%	\hspace*{-0.3cm}
%	\subfigure{\includegraphics[width=0.22\textwidth]{asymmetry/0.7,0.8,J=0.2}}
%	\caption{The quench evolution of the order paramters $\gamma_{1(2)}$ with (a) $\lambda_1=0.45,\lambda_2=0.55$ and $J_f=0.2$; (b) $\lambda_1=0.7,\lambda_2=0.8$ and $J_f=0.1$ with different initial state; (c)  $\lambda_1=0.7,\lambda_2=0.8$ and $J_f=0.2$ with different initial state; Other parameters are the same as in Fig.~\ref{table}.}
%	\label{asymmetric dynamic}
%\end{figure}

\begin{figure}[tb]
	\centering
	\includegraphics[width=0.42\textwidth]{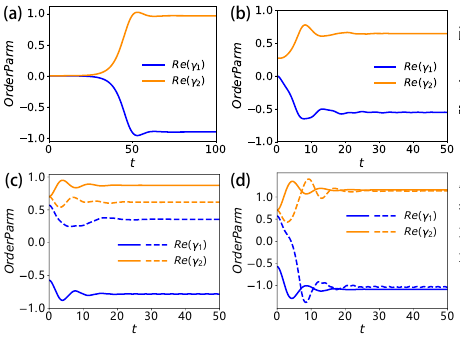}
	\caption{The quench evolution of the order paramters ${\rm Re}\gamma_{1(2)}$ with (a)$\lambda_1=0.25$, $\lambda_2=0.35$ and $J_f=0.4$; (b) $\lambda_1=0.45$, $\lambda_2=0.55$ and $J_f=0.2$; (c) $\lambda_1=0.7$, $\lambda_2=0.8$ and $J_f=0.1$ with different initial state; (d)  $\lambda_1=0.7$, $\lambda_2=0.8$ and $J_f=0.2$ with different initial state. Other parameters are the same as in Fig.~\ref{table}.}
	\label{asymmetric dynamic}
\end{figure}

The evolution of the order parameters ${\rm Re}\gamma_j$ is presented in Fig.~\ref{asymmetric dynamic}. In Fig.~\ref{asymmetric dynamic}(a), we consider a (NP\&NP) initial state with $\gamma_1=\gamma_2=0$. The final photon hopping rate exceeds $J_c$, hence the system reaches a (SRP\&SRP) final state. In Fig.~\ref{asymmetric dynamic}(b), a (NP\&SRP) initial state with $\gamma_1 \neq 0$ and $\gamma_2=0$ will reach a (SRP\&SRP) final state once we add a nonzero hopping interaction $J_f$. The solid lines in Fig.~\ref{asymmetric dynamic}(c) illustrate the evolution of a (SRP\&SRP) initial state. After the photon hopping is turned on, it evolves to a new (SRP\&SRP) final state. The dashed lines in Fig.~\ref{asymmetric dynamic}(c) illustrate the evolution of a different initial state for the same system. Here we see that it evolves to a different final state in comparison to the solid lines. This depicts the multistable region under the orange surface of Fig.~\ref{table}(b). In Fig.~\ref{asymmetric dynamic}(d), the solid and dashed lines again illustrate the evolution of two different initial states of the same system. Here they both evolve to the same final state. This corresponds to the situation of the SRP region above the orange surface of Fig.~\ref{table}(b). 

\begin{figure}[tbp]
	\centering
	\hspace*{-0.3cm}
	\includegraphics[width=0.5\textwidth]{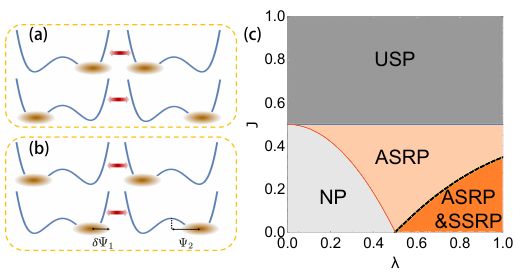}
	\caption{Schematic plot showing energy as a function of the order parameter for (a) ASRP and (b) SSRP, respectively. Each energy curve shows a double-well structure, reflecting the spontaneous breaking of the $Z_2$ symmetry for each cavity.  (c) The phase diagram with $\kappa=0.2$.}
	\label{Dicke_phase_diagram_analytical}
\end{figure}

\textit{Symmetric Dicke dimer.} 
In order to provide more insights, we now consider the special case where the two cavities are identical, i.e., all the parameters for the two cavities are the same. We are able to solve this system almost completely analytically by taking advantage of the symmetry between the cavities. Due to such symmetry, we can focus on two sets of solutions: the symmetric SRP (SSRP) and the anti-symmetric counterpart (ASRP) characterized by $\gamma_1=\pm\gamma_2$, $X_1=\pm X_2$, $Y_1=\pm Y_2$, and $Z_1=Z_2$, respectively. These two phases are represented in Fig.~\ref{Dicke_phase_diagram_analytical}(a-b) where we schematically show the energy as a function of the order parameter.  

Adding these conditions to the steady-state equation ${d\Psi_{\rm tot}}/{dt}=0$, we obtain (for details, see SM~\cite{supplementary}), apart from the trivial NP solution [$\gamma_j=0$ and $Z_j=-\frac{1}{2}$], the nontrivial solutions for spin-$z$ in SRP as 
\begin{equation}
    Z_{s(a)}^e=-({\omega_a}/{8\lambda^2})(\omega_c\pm2J+{\kappa^2}/{\omega_c})\,
\end{equation}
by requiring the determinant of the coefficient matrix ${\cal D}$ to vanish. Here the superscript ``$e$" stands for equilibrium steady state and the subscript ``$s(a)$" corresponds to SSRP (ASRP). In the following, we will apply the dimensionless parameters $\lambda\rightarrow\sqrt{\omega_c\omega_a}\lambda$, $\kappa\rightarrow\omega_c\kappa$, $J\rightarrow\omega_cJ$, and, for simplicity, set $\omega_c=\omega_a=1$.

Combining the equations of motion (\ref{eq:EOM}) and the spin conservation law $Z^2+X^2+Y^2=1/4$, the full set of the steady-state solutions can be found as~\cite{supplementary}
\begin{align}
	\label{ASRP_solution}
%	\begin{cases}
		{\rm Re}\gamma^e_{a(s)}&=\pm\frac{1}{4\lambda}\sqrt{(\frac{1}{4(Z^e_{a(s)})^2}-1)},\nonumber\\
		{\rm Im}\gamma^e_{a(s)}&=\kappa {\rm Re}\gamma^e_{a(s)},\nonumber\\
		X^e_{a(s)}&=\pm Z^e_{a(s)}\sqrt{(\frac{1}{4(Z^e_{a(s)})^2}-1)},\nonumber\\
		Y^e_{a(s)}&=0,\nonumber\\
		Z^e_{a(s)}&=\frac{1}{8\lambda^2}(\pm2J-1-\kappa^2),
%	\end{cases}
\end{align}
with $Z_{1a(s)}^e=Z_{2a(s)}^e\equiv Z^e_{a(s)}$, $X_{1a(s)}^e=\mp X_{2a(s)}^e\equiv X^e_{a(s)}$, $Y_{1a(s)}^e=\mp Y_{2a(s)}^e\equiv Y^e_{a(s)}$, $\gamma_{1a(s)}^e=\mp\gamma_{2a(s)}^e\equiv\gamma^e_{a(s)}$. 

To address the stability of these steady states, we take quantum fluctuations into account~\cite{PhysRevLett.122.193605} and define $\Psi_{\rm tot}=\Psi_{\rm tot}^e+\delta\Psi_{\rm tot}$. The dynamic equations governing the small fluctuations $\delta\Psi_{\rm tot}$ can be written compactly as ${d\delta\Psi_{\rm tot}}/{dt}=M(\Psi^e)\delta\Psi_{\rm tot}$, where the $8\times8$ matrix $M(\Psi^e)$ depends on the equilibrium solution $\Psi^e$, and thus takes different forms for different phases NP, ASRP and SSRP, whose specific expressions can be found in SM~\cite{supplementary}. 
%We have found the the $8\times8$ stable matrix can be effectively divided into four $4\times4$ blocks in all the three phases. The diagonal submatrixes are denoted as $\mathcal{A}_{4\times4}$, and counter-diagonal submatrixes are denoted as $\mathcal{B}_{4\times4}$. 
Remarkably, the parity symmetry of these phases will divide the dynamics into two invariant subspaces, which can be referred to as symmetric and anti-symmetric fluctuations, respectively. The total fluctuation thus take the form of
\begin{align}
	\delta\Psi_{\rm tot}(t)=&e^{M_St}\delta\Psi_S(0)\oplus e^{M_St}\delta\Psi_S(0)\notag\\
	&+e^{M_At}\delta\Psi_A(0)\oplus [-e^{M_At}\delta\Psi_A(0)].
\end{align}
Here, $M_{S(A)}$ is a $4\times4$ matrix dominated by the symmetric (anti-symmetric) quantum fluctuations $\delta\Psi_{{\rm tot} (S)}^T\equiv[\delta\Psi_S^T,\delta\Psi_S^T]$ ($\delta\Psi_{{\rm tot}(A)}^T\equiv[\delta\Psi_A^T,-\delta\Psi_A^T]$). For details, see SM~\cite{supplementary}.

The stable condition requires all the real parts of the eigenvalues for both matrices $M_S$ and $M_A$ to be negative~\cite{PhysRevA.35.5288}. Inserting the NP and ASRP solutions, the analytical phase boundary between them is obtained as \begin{equation}
    -4\lambda^2+1-2J+\kappa^2=0\,,
\end{equation} 
which is represented by the red solid line in Fig.~\ref{Dicke_phase_diagram_analytical}(c). Further analysis shows that this is a second order phase boundary~\cite{supplementary} as in the case of a single Dicke cavity. In the superradiance regime, we find that ASRP is always stable. By contrast, SSRP is only stable over part of the superradiance regime. The stability boundary of SSRP can be found as 
\begin{equation}
    J={(1+\kappa^2)}/{2}+16(Z^e_s)^3\lambda^2\,,
\end{equation}
which is represented by the black dashed line in Fig.~\ref{Dicke_phase_diagram_analytical}(c). The region below this line thus defines the multistable region where both SSRP and ASRP coexist.

The gray area in Fig.~\ref{Dicke_phase_diagram_analytical}(c) represents the unstable phase (USP) for all three quantum phases. This occurs when $J> (1+\kappa^2)/2$. For such large values of photon hopping rate, the system is energetically unbounded from below. An intuitive way to understand this instability is to simply focus on the photon hopping Hamiltonian $H_{XX}$. Using the Bogoliubov procedure, we can readily find that the eigen-energies of the photons are given by $\sqrt{\omega_c(\omega_c \pm 2J)}$. The lower branch becomes imaginary when $J>\omega_c/2$ [This condition becomes $J> (\omega_c^2+\kappa^2)/(2\omega_c)$ if we include cavity decay], signalling the instability of the system. Nonlinear effects not included in our model Hamiltonian, which are often present in real experimental systems~\cite{PhysRevLett.115.180404}, may be able to provide a mechanism to stabilize this region. These nonlinear effects are often weak and negligible far away from the instability boundary. However, when we approach the boundary, the nonlinear energy scale becomes comparable to the photonic eigen-energy $\sqrt{\omega_c(\omega_c\pm2J)}$ and its effect can no longer be neglected.   
%It also guarantees the system potential bounded in large $J$. Thus, we are supposed to modify the Hamiltonian by adding the nonlinear effect into the original Hamiltonian. , resulting from an unbounded downward harmonic oscillator potential
As a concrete example, we include the Kerr nonlinear term, $\sum_{j}\chi a_j^\dagger a_j^\dagger a_ja_j$, which describes the two-photon interaction in the cavity. The modified Hamiltonian reads $H=H_{\rm sys}+\sum_{j}\chi a_j^\dagger a_j^\dagger a_ja_j$. With a detailed numerical calculation shown in SM~\cite{supplementary}, it turns out that the phase diagram with $\chi\neq0$ still matches well with the $\chi=0$ case [Fig.~\ref{Dicke_phase_diagram_analytical}(c)] when the hopping rate $J$ is not that large, while the region in USP are stabilized and become part of the superradiance region (see Fig.~S1 (d) and Fig.~S3~\cite{supplementary}).

\begin{figure}[tb]
	\centering
	\hspace*{-0.4cm}
	\includegraphics[width=0.52\textwidth]{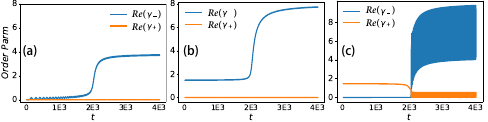}
	\caption{The adiabatic evolution of a closed system. (a) $\lambda=0.4$ and $J_i=J_c$ located at the NP-ASRP boundary, (b) $\lambda=0.8$, $J_i=0$, and the initial state being ASRP, (c) $\lambda=0.8$, $J_i=0$, and the initial state being SSRP. In all plots, $J_f=0.4$ is fixed. Here ${\rm Re}\gamma_\pm = {\rm Re}\gamma_1 \pm {\rm Re}\gamma_2$ can be regarded as the order parameter for SSRP and ASRP, respectively.}
	\label{Dicke dimer dynamic}
\end{figure}

%Similar to the discussion in the generalized Dicke dimer, we investigate the quench process for this symmetric case, where a sudden change $J_i=0\rightarrow J_f=0.4$ drives the evolution. The initial NP in Fig.~\ref{Dicke dimer dynamic} (a) can only be transferred to the ASRP state once the hopping rate $J$ goes across the NP-ASRP boundary. Such boundary is attributed to the instability of anti-symmetric quantum fluctuation, so the final stable state tends to be in odd parity, i.e., the ASRP state. As is shown in Fig.~\ref{Dicke dimer dynamic} (b), the initial ASRP state won't go through a phase transition, with the symmetry of the state unchanged. 
%Similarly, the initial SSRP will evolute into a new SSRP as $J$ is below the 1st order critical point between the ASRP and SSRP[see in subfigure (c)]. Although the ASRP state is also supported in this region, the SSRP can be approximated bounded in the symmetric subspace, as long as that the quantum fluctuation is not that large. 
%While in Fig.~\ref{Dicke dimer dynamic} (c), the even parity of SSRP will be broken as $J$ goes beyond the critical point, where a new balance is established due to the dissipation, and thus it is transferred to ASRP.

\textit{Ground state in closed system.} Using the method of Holstein-Primakoff mapping~\cite{supplementary}, the ground state of the closed system (i.e., without cavity decay, $\kappa_j=0$) can be determined. The ground-state phase diagram is similar to the steady-state phase diagram for the open system we presented here.  For $J=0$, both ASRP and SSRP are degenerate and represent the ground states of the system in the superradiance regime. For finite $J$, ASRP always possesses lower energy than SSRP. The SSRP could exist as a metastable state that is dynamically stable in the counterpart of the multistable region. 

%In another region SSRP occurs, the system will be in a unstable equilibrium state, which means that the quantum fluctuation will cause the fast dissipation of the SSRP state.

In Fig.~\ref{Dicke dimer dynamic}, we illustrate 
the adiabatic evolution of the closed system with two identical cavities when the photon hopping rate $J$ is slowly ramped. Specifically, we choose a slowly changed hopping rate $J(t)=[({J_f-J_i})/{\pi}]\text{atan}[k(t-t_f/2)]+({J_f+J_i})/{2}$
%\begin{align}
%	J(t)=\dfrac{J_f-J_c}{\pi}atan[k(t-t_f/2)]+\dfrac{J_f+J_c}{2},
%\end{align}
with $kt_f\gg1$. Here, we set $k=0.02$ and $t_f=4000$.
%\begin{figure}[tb]
%	\centering
%	\subfigure{\includegraphics[width=0.157\textwidth]{order param/adiabatic_lambdax=0.4,J=0.4}}
%	\subfigure{\includegraphics[width=0.157\textwidth]{order param/adiabatic_lambdax=0.8,J=0.4(asym)}}
%	\subfigure{\includegraphics[width=0.157\textwidth]{order param/adiabatic_lambdax=0.8,J=0.4(sym)}}
%	\caption{The adiabatic evolution for the several cases (a)$\lambda_x=0.4,J_f=0.4$, and $J_i=J_c$ is the critical point between the NP and ASRP; (b)$\lambda_x=0.8,J_f=0.4,J_i=0$, and the initial state is the ASRP; (c)$\lambda_x=0.8,J_f=0.4,J_i=0$, and the initial state is the SSRP.}
%	\label{Dicke dimer adiabatic evolution}
%\end{figure}
In Fig.~\ref{Dicke dimer dynamic}(a), we choose the hopping rate changes from the critical boundary $J_c$ of the NP stable region. It is shown that the initial NP gradually transferred into an ASRP state as the hopping rate increases. In Fig.~\ref{Dicke dimer dynamic}(b), the initial ASRP gradually transfer into a new stable ASRP. In both cases, the order parameter follows closely the change of $J(t)$. However, the initial SSRP cannot transfer smoothly into a stable ASRP if $J_f$ exceeds the SSRP stability boundary, as shown in Fig.~\ref{Dicke dimer dynamic}(c), demonstrating the dynamical instability of the SSRP for large $J$. That an initial SSRP cannot evolve into an ASRP state is because the adiabatic evolution of the closed system maintains the parity symmetry.
%And the order parameter will suffer a sudden change at the critical boundary, and even the adiabatic added hopping interaction fail to transfer between the two SRPs with different parity. 
In contrast, the cavity decay in an open system breaks the $Z_2$ symmetry, and allow the system to relax into the new stable SRP state regardless the symmetry of the initial state.

\textit{Experimental feasibility.} Towards the realistic implementation, here we propose a potential experimental protocol in a circuit quantum electrodynamics system, which, as one of the most promising platform for quantum computing~\cite{RN69}, has also shown feasibility to achieve ultrastrong coupling between artificial atoms and optical modes, thus providing capabilities to simulate the light-matter coupled phase transitions~\cite{PhysRevLett.112.173601,PhysRevLett.113.093602,RN84,RN86,RN87,RN88}. A detailed description of the proposal can be found in the SM~\cite{supplementary}.

\textit{Summary.} In order to explore how two different quantum phases affect each other, we investigate a Dicke dimer consisting two Dicke cavities with photon hopping between them. We show that photon hopping plays a crucial role for this system. Significantly, for the generic dimer system where the two cavities are not necessarily identical, the existence of all the possible phases is analytically proved. According to our results, the hopping term makes the Dicke dimer favor SRP (i.e., lowers the critical atom-light coupling strength to reach superradiance) in both cavities, providing potential applications to manipulate quantum phases by introducing couplings between cavities. In addition, a multistable region occurs in the superradiance region. As a special case convenient for analytical study, we give the phase diagram for symmetric Dicke dimers. In order to study the multistable region where both SSRP and ASRP occur, the ground state in closed system is also obtained. And by introducing two-photon interactions, the unstable region in the original phase diagram is analyzed, resulting in convergent solutions numerically. Our work explores the exotic physical phenomena on the ``contact" between two quantum phases, analogous to the thermal contact between two materials in classical phase transitions, which paves the way to investigating the interaction between different quantum phases.

\begin{acknowledgments}
This work is supported by the National Natural Science Foundation of China (Grant Nos. 12125402, 12074428, 92265208), the National Key R\&D Program (Grant No. 2022YFA1405300), and the Innovation Program for Quantum Science and Technology (No. 2021ZD0301500). F. S. acknowledges the China Postdoctoral Science Foundation (Grant No. 2020M680186). H.P. is supported by the US NSF PHY-2207283 and the Welch Foundation (Grant No. C-1669).
\end{acknowledgments}

\bibliography{ref}

\begin{thebibliography}{32}
\bibitem [{Z2_()}]{Z2_symmetry}%
  \BibitemOpen
  \href@noop {} {}\bibinfo {note} {The Dicke Hamiltonian possesses the $Z_2$
  symmetry. In the absence of the photon hopping ($J=0$), the parity operators
  associated with each cavity $\hat{P}_j=\exp(i\pi \hat{N}_j)$, where
  $\hat{N}_j\equiv a_j^\dagger a_j+S_j^z+N/2$, are conserved, corresponding to
  the $Z_2 \times Z_2$ symmetry. In the presence of the photon hopping ($J\neq
  0$), Only the total parity operator
  $\hat{P}=\exp(i\pi\sum_{j=1}^{2}\hat{N}_j)$ is conserved, defining a total
  $Z_2$ symmetry for the dimer system.}\BibitemShut {Stop}%
\end{thebibliography}
	
\end{document}